\documentclass[twocolumn,showpacs,preprintnumbers,amsmath,amssymb,nofootinbib]{revtex4}

\usepackage{graphicx}
\usepackage{dcolumn}
\usepackage{bm}
\usepackage{graphics}
\usepackage{textcomp}
\usepackage{amssymb}

\usepackage{times}

\newcommand{\be}{\begin{equation}}
\newcommand{\ee}{\end{equation}}
\newcommand{\bea}{\begin{eqnarray}}
\newcommand{\eea}{\end{eqnarray}}

\newcommand{\eqintext}[1]{\mbox{$#1$}}

\begin{document}

\title{Non-Gaussian Velocity Distributions in Optical Lattices}

\author{Johan Jersblad$^{1,2}$}
\author{Harald Ellmann$^{1}$}
\author{Kristian St{\o}chkel$^{2}$}
\author{Anders Kastberg$^{2}$}
\author{Laurent Sanchez-Palencia$^{3}$}
\author{Robin Kaiser$^{4}$}
\affiliation{
   $^{1}$Department of Physics, Stockholm University, S-106 91
   Stockholm, Sweden\\
   $^{2}$Department of Physics, Ume{\aa} University,
   S-901 87 Ume{\aa}, Sweden\\
   $^{3}$Laboratoire Kastler-Brossel, D{\'e}partement
   de Physique de l'Ecole Normale Sup{\'e}rieure, 24 rue Lhomond, F-75231
   Paris cedex 05, France\\
   $^{4}$Laboratoire Ondes et D{\'e}sordre, FRE 2302, 1361 route des Lucioles,
   F-06560 Valbonne, France
}

\date{7 September 2003}

\begin{abstract}
We present a detailed experimental study of the velocity distribution
of atoms cooled in an optical lattice. Our results are supported by
full-quantum numerical simulations. Even though the Sisyphus effect,
the responsible cooling mechanism, has been used extensively in many
cold atom experiments, no detailed study of the velocity distribution
has been reported previously. For the experimental as well as for the numerical 
investigation, it turns out that a Gaussian function is not the one that best 
reproduce the data for all
parameters. We also fit the data to alternative functions, such as
Lorentzians, Tsallis functions and double Gaussians. In particular, a
double Gaussian provides a more precise fitting to our results.
\end{abstract}


\maketitle

\section{Introduction}
\label{sec:intro}
Laser cooling is now a well established technique to produce narrow
velocity distributions for dilute samples of atomic gases (see e.g.
\cite{nobel97}). The interaction between the atoms and the radiation
modes removes kinetic energy from the atoms, and extremely cold
samples can be obtained. In the standard context of Doppler or
sub-Doppler laser cooling, atom-atom interactions are neglected and
hence a thermodynamic temperature cannot be defined. Nevertheless,
measured velocity distributions are generally very well fitted by a
Gaussian function, and assigning a \mbox{\textquoteleft kinetic
temperature\textquoteright} to the distribution is a useful way to
characterize a laser cooled atomic sample.

One of the simplest theoretical models of laser cooling assumes a
moving two-level atom interacting with counter-propagating pairs of
laser beams, tuned slightly below the atomic resonance (Doppler
cooling \cite{doppler}). This will yield Doppler shifts, asymmetric
with regards to velocity, and thus a damping force (friction). 
Doppler cooling is counteracted by momentum diffusion due to
absorption and emission of photons. If a spatial average is taken of
diffusion as well as friction, one obtains a stationary Gaussian
velocity distribution. This is valid since, in steady-state, most
atoms have velocities well above spatial modulations in the light
shift potential (caused by the interaction between the induced dipole
moment and the light), and thus the dynamics can be described in terms
of a Fokker-Planck equation with constant friction and diffusion
coefficients. High irradiance results in light shifts of the involved
energy levels that can be comparable to the kinetic energy, and one
can no longer assume a constant velocity as atoms travel over a
wavelength. Spatial averaging can still be performed, but one does
not obtain the standard description of laser cooling in terms of
competition between a friction force and a diffusion effect, since
these are not simply functions of velocity. The resulting velocity
distribution will in this case not be Gaussian and different
distributions have been proposed \cite{Kazantsev}. However, for
practical Doppler cooling configurations, this effect is negligible,
and there are no known observations of clearly non-Gaussian
distributions.

For a multilevel atom, population transfer and coherences between
degenerate levels open up the possibility for more subtle cooling
mechanisms. These are not limited by the radiative lifetimes of the
upper levels, and can therefore lead to narrower distributions. In
particular, Sisyphus cooling \cite{Dalibard89, Chu891, Chu892,
Phillips89} is based on a laser beam configuration that results in a
periodic modulation of the polarization of the light, and thus
spatially modulated optical pumping and steady-state population
distribution between different degenerate substates. The light shift
will also be periodic, and will differ for different substates. The
combination of hamiltonian motion and optical pumping cycles transfers
atomic energy to the vacuum modes \cite{Dalibard89,Chu891,cohen90}. A
rule of the thumb for Sisyphus cooling tells us that the
\mbox{\textquoteleft temperatures\textquoteright} obtained correspond
to kinetic energies that are of the order of the light shift. This
behavior has been experimentally verified \cite{Salomon90, Gatzke,
Jersblad, Ellmann} down to kinetic temperatures of a few recoil
energies. A seminal analysis of Sisyphus cooling, by Dalibard and
Cohen-Tannoudji \cite{Dalibard89}, is again based on spatially
averaged friction and diffusion coefficients. Even though the final
regime corresponds to a situation where one can no longer assume atoms
moving at constant velocity over many wavelengths, the scaling law
obtained by this approach appears to be excellent.

In more rigorous full quantum mechanical analyses, Castin et al. 
\cite{Castin,Castin91} find that Sisyphus cooling ought to lead to
non-Gaussian distributions. In particular, for irradiances close to the lower
limit for efficient laser cooling, the effects of recoils due to absorbed and emitted
photons become prominent. Then, atomic trajectories become very irregular and
the velocity cannot be assumed to be constant. Therefore one cannot compute a spatially averaged velocity dependent force. Also, the atoms will be trapped in
microscopic potential minima (forming optical lattices
\cite{mennerat02,jessen}), and the ensemble should be characterized by
a distribution of vibrational modes and unbound modes, rather than by
a velocity distribution.

Essentially all experimental investigations of Sisyphus cooling result
in distributions that are well fitted by Gaussians. The reason for
this is probably a combination of several facts. Many experiments are
done in a regime where an average friction coefficient seems adequate
(sufficiently large light shift). The deviations from Gaussian
distributions are subtle and are mainly hidden in the noisy wings of
the recorded distribution. Furthermore, it is difficult to set-up an
experimental velocity probe with the required resolution. 
Nevertheless, deviations from Gaussian velocity distributions for
laser cooled atoms have been reported in one recent paper
\cite{Salomon}. However, to our knowledge, there has been no
systematic experimental study of the non-Gaussian distributions,
nor any attempts to approach the observed distributions with more
precise functions.

In this work, we report a detailed study of velocity distributions, as
a function of the irradiance (and thus the light shift) for a three
dimensional Sisyphus cooling configuration. We also perform a
one-dimensional numerical simulation of velocity distributions, based
on a full-quantum Monte-Carlo wave function technique. This is
applied for the atomic angular momentum which is relevant in our
experiment. We fit the recorded data, the experimental as well
as the numerical, to different functions and compare the outcomes.

\section{Fitting functions and motivations}
\label{Fitting functions and motivations}
The main purpose of this paper is to present more details about the
velocity distributions of atomic samples cooled and trapped in optical
lattices, where the Sisyphus cooling theory is expected to apply. A
further step is to provide a function that gives a good approximation
of the velocity distribution. The choice of a fitting function is
made difficult by the complex dynamics of the atoms in the lattice. 
Indeed, even if the seminal process described in \cite{Dalibard89}
gives very good insights in the dynamical behavior of the atoms, it is
not sufficient in regimes relevant for typical experimental
situations, where the intercombination of hamiltonian motion in the
modulated potentials and optical pumping cycles, with time scales of
the same order, makes it difficult to perform analytical calculations
\cite{Castin}. Along the following lines we justify {\it a priori}
the choice of three types of functions (Gaussian, Tsallis and double
Gaussian) that we used to fit the experimental and the numerical
recorded data. As we will see, these choices are based on simple
considerations about well-known generalizations of the model presented
in ~\cite{Dalibard89}.

\paragraph{Gaussian function}
In the standard description of 1D-Sisyphus cooling, the internal
atomic state is adiabatically eliminated in such a way that the atomic
dynamics is described in simple terms of a force
\eqintext{F\left(v\right)} and fluctuating forces of momentum
diffusion coefficient \eqintext{ D_{v}\left(v\right)}. 
\eqintext{F\left(v\right)} accounts for the optical pumping-assisted
Sisyphus cycles and \eqintext{ D_{v}\left(v\right)} corresponds, on
the one hand, to the random recoils due to absorption and emission of
photons, and on the other hand, to changes of potential curves. The
velocity distribution, $W(v)$, is thus governed by a Fokker-Planck
equation (FPE) \cite{Risken89,VanKampen92}:
\begin{equation}
\partial_{t} W = -\partial_{v}  \left( \frac{1}{M} F\left(v\right) W \right) +  
\partial_{v}  \left(  D_{v}\left(v\right) \partial_{v}  W \right);
\label{eq:fpe}
\end{equation}
with $M$ being the atomic mass. In the linear regime for the atomic velocity, one finds \cite{cohen90}:
\bea
& & F\left(v\right) = -\alpha v \\
& &  D_{v} =  D_{v}^{(1)} +  D_{v}^{(2)}. \nonumber
\label{eq:sisyphuslinear}
\eea
In this context, \eqintext{\alpha} and \eqintext{ D_{v}} depend on the
lattice parameters and are independent of
the velocity. \eqintext{ D_{v}^{(1)}} corresponds to the random absorption  
and emission of photons while
\eqintext{ D_{v}^{(2)}} represents the fluctuations of the light-shift  
induced force \cite{cohen90}. The steady-state solution
of Eq.~(\ref{eq:fpe}) with vanishing probability current  
(\eqintext{-F\left(v\right)W +  MD_{v}
\left(v\right) \partial_{v} W = 0}) is thus a Gaussian function with  
rms width
\eqintext{\sigma_{v}=\sqrt{ MD_{v}/\alpha}}:
\begin{equation}
W\left(v\right) = W_{0} \exp{\left(-\frac{\alpha v^2}{2  MD_{v}}\right)}.
\label{eq:gaussian}
\end{equation}

 \paragraph{Tsallis function}
 Beyond the linear regime for atomic velocity, the friction force and  
 the velocity diffusion coefficients have to
 be refined  into \cite{Castin,Hodapp95}:
 \bea
 & & F\left(v\right) = \frac{-\alpha  
 v}{1+\left(v/v_{\textrm{c}}\right)^2} \label{eq:sisyphus_nonlinear} \\
 & & D_v\left(v\right) = D_v^{(1)} +  
 \frac{D_v^{(2)}}{1+\left(v/v_{\textrm{c}}\right)^2}, \nonumber
 \eea
 where \eqintext{v_{\textrm{c}}} is the capture velocity which  
 corresponds to the typical atomic velocity above which
 the Sisyphus process breaks down.
 Now, it is straightforward to show that the steady-state solution with  
 vanishing probability current of
 Eq.~(\ref{eq:fpe}) reads \cite{Lutz}
 \begin{equation}
 W(v) = W_0 \left[1-\beta\left(1-q\right) v^2\right]^\frac{1}{1-q}
 \label{eq:tsallis}
 \end{equation}
 \bea
 q = 1+\frac{2 MD_v^{(1)}}{\alpha v_{\textrm{c}}^2} & \textrm{and} &  
 \beta = \frac{\alpha / 2 M}{D_v^{(1)}+D_v^{(2)}}.
 \label{eq:tsallis_coeff}
 \eea
 The function in Eq. (\ref{eq:tsallis}) is the so-called Tsallis
 function and is in fact very general. It particularly provides a
 broad class of fitting functions including Gaussian functions
 (\eqintext{q} approaching one), Lorentzian functions (\eqintext{q=2})
 and inverted parabolas (\eqintext{q=0}). At this stage, it is
 interesting to note that the Tsallis function has been introduced in
 the context of non-extensive thermodynamics
 \cite{Tsallis1,Tsallis1bis}. The large amount of literature in this
 context allows one to find many papers dealing with problems already
 addressed in laser cooling; in particular anomalous diffusion in the
 presence of external forces \cite{Quarati,Tsallis2,Plastino95},
 multiplicative noise problems, and the relation to the edge of chaos
 in mixed phase space dynamics \cite{Lyra98,Weinstein02}. It is
 known that Sisyphus cooling can give rise to anomalous
 diffusion \cite{Katori97, Zoller96}, in particular for shallow
 optical potentials, where an atom can travel over many wavelengths
 before being trapped again.
Even though we do not have a detailed analysis of the dynamics of the atoms in an optical lattice, for parameters corresponding to our situation, one cannot rule out anomalous diffusion and/or chaotic behavior.

 \paragraph{Double Gaussian function}
 As Sisyphus cooling results in a situation where the kinetic energies
 of the atoms are of the order of the light shift potential, one can
 neither neglect atoms with lower energy ({\textquoteleft trapped\textquoteright} in 
 the potential wells) nor those moving more or less freely above the potential
 modulation (as in a {\textquoteleft conduction band\textquoteright}). This leads to 
 a description of the atomic sample in terms of a bimodal dynamics. Note that such
 a bimodal description has been shown to be relevant for the
 prediction of the diffusive properties of atoms in an optical lattice
 \cite{palencia}. The kinetic equation of the {\textquoteleft high 
 energy\textquoteright} atoms might very well be described by spatially averaged 
 friction and diffusion coefficients resulting in a Gaussian distribution as shown
 previously. The {\textquoteleft low energy\textquoteright} atoms will be trapped, 
 and subject to a different kinetic equation, and we assume that their velocity
 distribution is again a Gaussian. Our trial function is thus the sum
 of two Gaussian distributions with different widths (double Gaussian). One 
 corresponding to {\textquoteleft trapped\textquoteright} atoms and the other 
 one to {\textquoteleft high energy\textquoteright} atoms.

\section{Experiments}
\subsection{Experimental setup}
\begin{figure}[h]
\begin{center}
\includegraphics[scale=0.5]{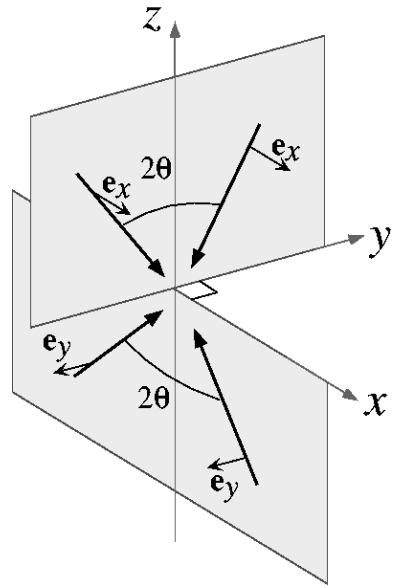}
\caption{Beam configuration of the 3D lin $\perp$ lin optical
lattice. Two beam pairs propagate in the $xz$- and $yz$-planes, and  
are orthogonally polarized along the $y$- and $x$-axes
respectively. They form an angle of $\theta=45^{\circ}$ with the
$z$-axis. \label{beamconfig}}
\end{center}
\end{figure}

The experimental setup has been described in detail previously (see
e.g. \cite{Jersblad, Ellmann}). Briefly, we first accumulate
\eqintext{^{133}}Cs atoms in a magneto-optic trap (MOT). We adjust
the irradiance and the detuning, then we turn off the magnetic field
and leave the atoms in an optical molasses with even further reduced
irradiance. Thus we cool the atoms to a temperature of 3-4 $\mu$K.
The atoms are transfered to a three-dimensional optical lattice, which
is based on four laser beams of equal irradiance and detuning (for a
review of optical lattice set-ups, see e.g. \cite{mennerat02} or
\cite{jessen}). The detuning is a few tens of $\Gamma$ below the
($F_{\textrm{g}} = 4 \rightarrow F_{\textrm{e}}$ = 5) resonance for
the \eqintext{^{133}}Cs D2 line at 852 nm (\eqintext{\Gamma=2\pi \cdot 5.21} MHz is the
linewidth of the excited state). The detuning (\eqintext{\Delta}) and
irradiance (\eqintext{I}) of the beams can be easily changed in order
to control the depth of the light shift potential \eqintext{U_0
\propto I/|\Delta|}. The beams are aligned as in Fig. 
\ref{beamconfig}: two laser beams are linearly polarized along the
$x$-axis and propagate in the $yz$-plane symmetrically with respect to
the $z$-axis, whereas the other two beams are polarized along the
$y$-axis and propagate in the $xz$-plane symmetrically with respect to
$z$. This yields a tetragonal pattern of points with pure circular
polarization, alternately $\sigma^+$ and $\sigma^-$. These points
correspond to potential wells where the atoms are trapped and
optically pumped into the extreme $m_{F}$-levels (+4 and -4
respectively in $\sigma^+$- and $\sigma^-$-wells).

For high atomic velocities, this configuration will correspond to a
three-dimensional version of the Sisyphus cooling model. As the atoms
approach equilibrium, their kinetic energies will get lower than the
modulation depth of the optical potential, and thus atoms become
trapped in lattice sites. They will get distributed in bound states,
where the lowest states closely resemble harmonic oscillator states.

In two different sets of runs, we let the atoms equilibrate in the optical lattice 
for 25 ms and 50 ms respectively. The velocity distribution is then recorded 
with a standard time-of-flight method (TOF) \cite{Phillips89}.  After the lattice 
period the trapping field is turned off, and the atoms are released in the 
gravitational field; approximately 5 cm below the trap region a thin sheet of
resonant laser light crosses the vertical axis along which the atoms
fall, and the induced fluorescence is recorded with a photo-diode. 
Each vertical velocity component at the time of release will
correspond to a specific arrival time at the probe beam. The probe
beam is carefully spatially filtered and focussed by a cylindrical
lens. The interaction region is less than 50 $\mu$m thick, and the
trapped cloud of atoms is approximately 400 $\mu$m in diameter. This
gives a velocity resolution of 0.05 mm/s, or 0.015 $v_{\textrm{R}}$
(where \mbox{$v_{\textrm{R}}$ = 3.5 mm/s} is the velocity
corresponding to the recoil from one absorbed photon resonant with the
D2-line). Our statistics is good enough not to contribute to this
resolution. The optical lattice beams are turned off, by switching an
acousto-optic modulator, faster than a microsecond. This is fast
enough to avoid adiabatic release of the atoms in the lattice, which
could greatly influence the velocity distribution, in particular in
the high velocity tails \footnote{ If the optical lattice beams are
turned off too slowly, the atoms may partially equilibrate in the
gradually decreasing potential. There may also be adiabatic cooling
\cite{Kastberg}. In both these cases, the cooling during a slow turn
off can greatly influence the velocity distribution, in particular in
the high velocity tails. Such adiabatic switching is often used in
order to achieve lower {\textquoteleft temperatures\textquoteright}.}.

\subsection{Experimental results}
We recorded the velocity distributions for several modulation depths
and we fitted them with the functions introduced in
section~\ref{Fitting functions and motivations} with a slight
modification that accounts for atomic losses. During the long
optical lattice phase, we have a constant loss of atoms, probably
due to spatial diffusion. Therefore, the baseline is higher for
atoms with a downward velocity (short times, $v<0$) than it is for
atoms with a upward one ($v>0$). We compensate for this by adding a
sharp step function to the fit, with the amplitude of the step as a
free parameter. The amplitude of this step function is found to
increase sharply for decreasing potential depths between
$U_{\textrm{0}} = 200 E_{\textrm{R}}$ and $100 E_{\textrm{R}}$. A
probable reason is that spatial diffusion increases rapidly when the
potential depth falls below some threshold, which takes place for
higher potential depths than the threshold for cooling (usually
called {\textquoteleft d\'ecrochage\textquoteright}) \cite{Hodapp95}. This 
is consistent with previous studies \cite{carminati}. In principle, we could have used a linearly decreasing function instead of the step function, but then this would have had be terminated by a sharp step. We avoied this in order to minimize the number of free parameters and also because we wanted to simplify as much as possible in the absence of detailed knowledge of the loss of atoms.

\begin{figure}[h]
\begin{center}
\includegraphics[scale=0.45]{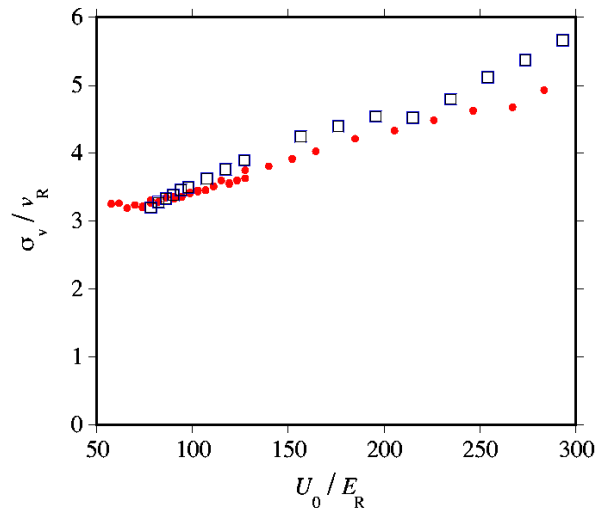}
\caption{(Color online) The rms width $(\sigma_{v})$ of the measured velocity distributions
(filled circles) as a function of the modulation depth of the
potential. Also shown is numerically simulated data (open squares)
in the same range (c.f. chapter~\ref{simulations}).
\label{velspread}}
\end{center}
\end{figure}

In Fig.~\ref{velspread}, we show the rms width of the distributions,
$\sigma_{v}$, as a function of the depth of the optical potential
$U_0$, as derived from the fits to single Gaussian functions. The
width, which is normally associated with a kinetic temperature,
increases for deeper potential depths as usual.

\begin{figure}[h]
\begin{center}
\includegraphics[scale=0.60]{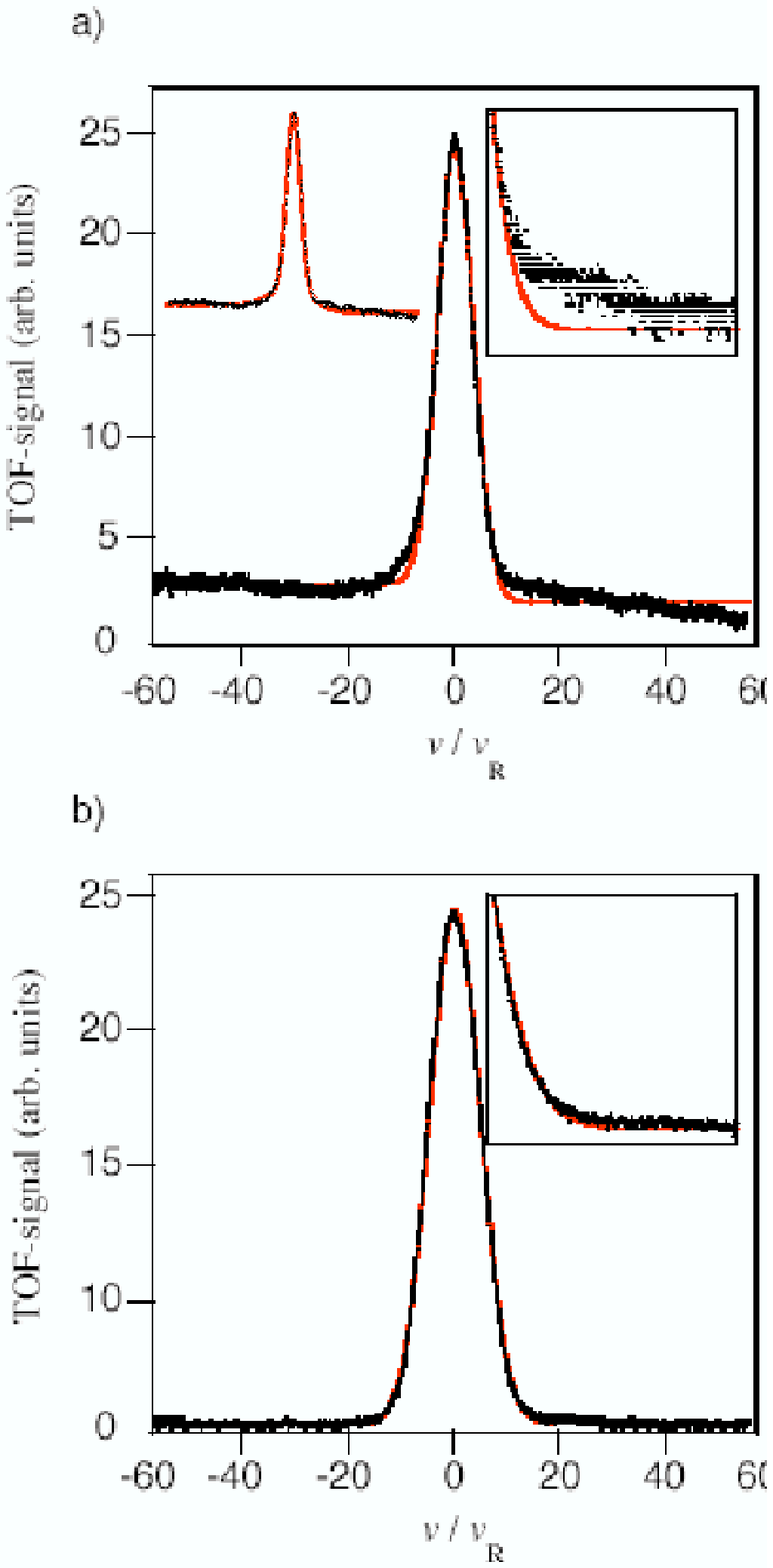}
\caption{(Color online) Experimentally recorded velocity distributions with fits to simple
Gaussians. This data is recorded with an equilibration time of 25 ms. For a the modulation depth of the optical potential was $U_{\textrm{0}} = 78 E_{\textrm{R}}$ and the
shown data is an average of 20 TOF measurements. For b the corresponding facts were $U_{\textrm{0}} = 285 E_{\textrm{R}}$ and an average of 5 TOF measurements. The insets in the top right corners show magnifications of portions of the wings of the distributions. The inset in the top left corner of a show the same data with a fit to a double gaussian.
\label{veldistr25}}
\end{center}
\end{figure}

\begin{figure}[h]
\begin{center}
\includegraphics[scale=0.60
]{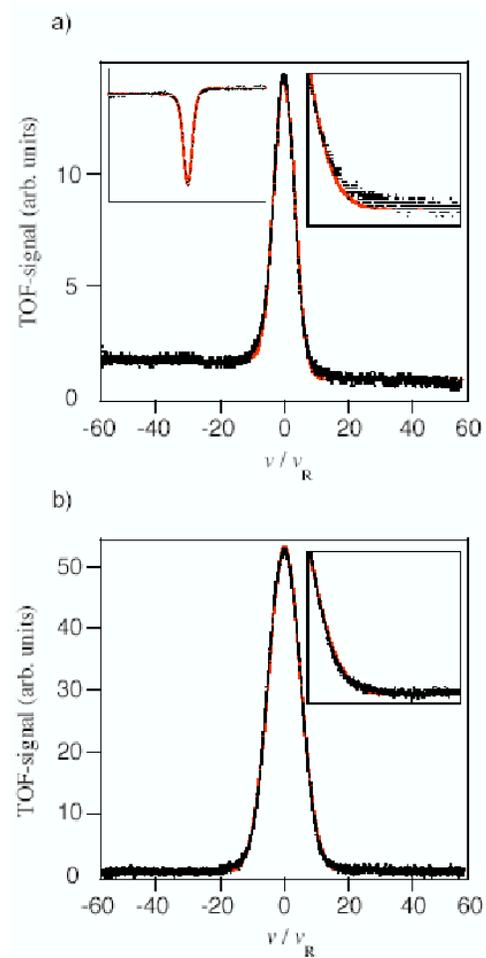}
\caption{(Color online) Experimentally recorded velocity distributions with fits to simple
Gaussians. This data is recorded with an equilibration time of 50 ms. For a the modulation depth of the optical potential was $U_{\textrm{0}} = 78 E_{\textrm{R}}$ and the
shown data is an average of 20 TOF measurements. For b the corresponding facts were $U_{\textrm{0}} = 285 E_{\textrm{R}}$ and an average of 5 TOF measurements. The insets in the top right corners show magnifications of portions of the wings of the distributions. The inset in the top left corner of a show the same data with a fit to a double gaussian.
\label{veldistr50}}
\end{center}
\end{figure}

In Figs.~\ref{veldistr25} and \ref{veldistr50}, typical recorded velocity distributions, together 
with Gaussian fits, are shown for low and high modulation depths. Figure \ref{veldistr25} shows data taken with an equilibration time of 25 ms, and for Fig. \ref{veldistr50} the equilibration time was 50 ms. This corresponds to typically $10^{6}$ radiative lifetimes. The plots with low irradiance 
are averages of twenty measurements and those of high irradiance of 
five measurements. For high values of the irradiance, a Gaussian function
fits the velocity distribution extremely well. However, for low irradiance, it
is clear that the wings of the distribution is not so well fitted. For the short
equilibration time, this is more pronounced.


\begin{figure}[h!]
\begin{center}
\includegraphics[scale=0.45]{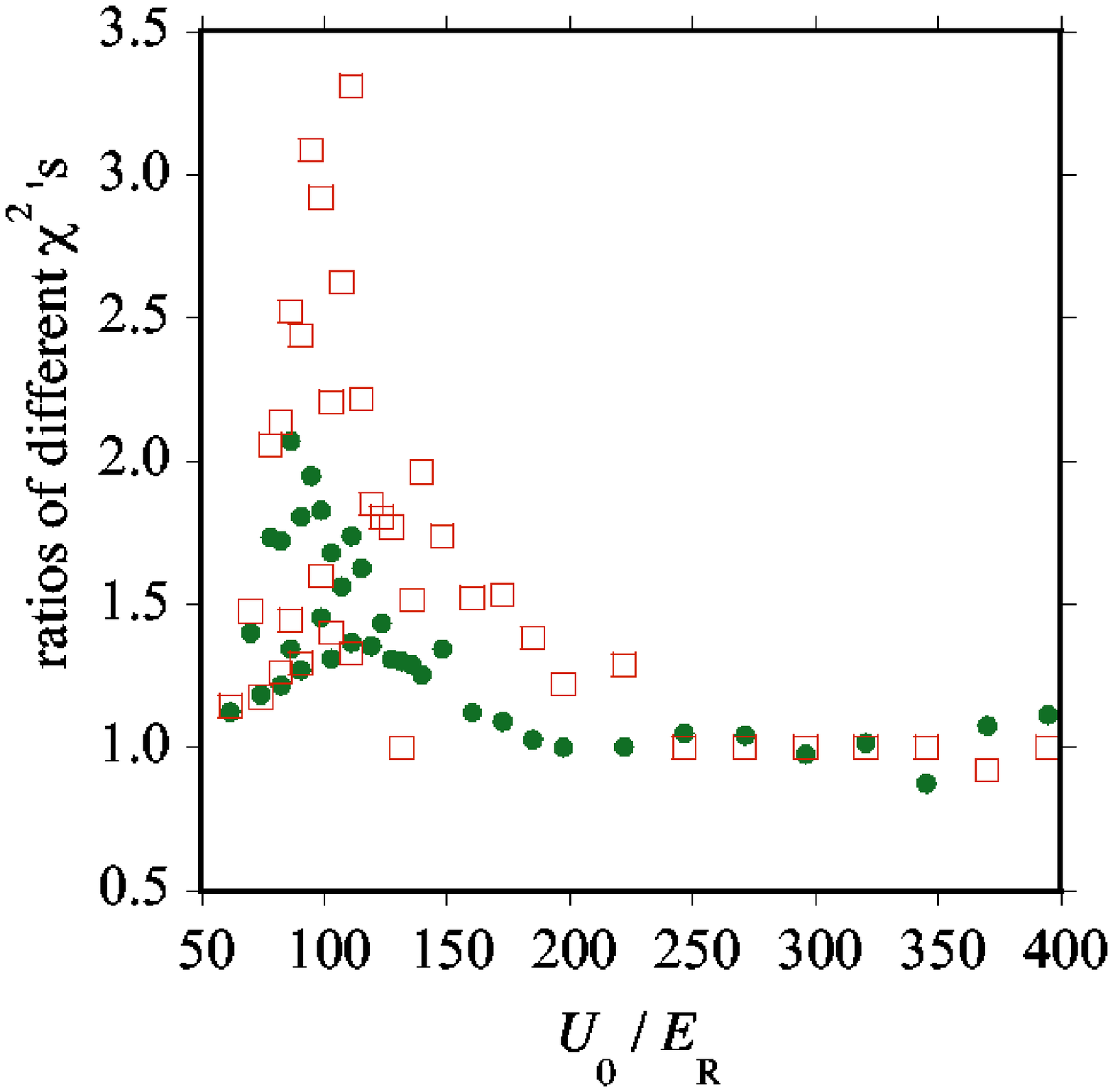}
\caption{(Color online) Comparisons between different fits of the measured distribution
for 25 ms equilibration time shown as ratios between 
unnormalized values of $\chi^2$ as a function of modulation
depth of the potential. The circles are 
$\chi^2_{\textrm{1Gauss}}/\chi^2_{\textrm{Tsallis}}$
and the squares are $\chi^2_{\textrm{1Gauss}}/\chi^2_{\textrm{2Gauss}}$.
\label{chisq}}
\end{center}
\end{figure}

For all data, even below {\textquoteleft d\'ecrochage\textquoteright}, the attempt with 
Lorentzian fits worked very poorly. Fits to double Gaussians and Tsallis
functions, however, reproduced recorded distributions better than
single Gaussians. In insets in figs. \ref{veldistr25}a and \ref{veldistr50}b we show fits to double Gaussians for shallow potentials. In Fig. \ref{chisq}, we compare the errors from
the fits for these three types of functions. When the irradiance is varied,
the signal-to-noise changes substantially, and so does the magnitude of 
the loss pedestal at short times, and the width and shape 
of the distribution. This makes it very hard to achieve a consistent normalization 
of the quality of the fits. The value of $\chi^2$ ($\chi^2 = \Sigma (y_{\textrm{i}}-x_{\textrm{i}})^2$, 
where $y_{\textrm{i}}$ is the measured and $x_{\textrm{i}}$ the fitted value) for an individual fit 
includes information about both noise and systematic deviation from the 
fit function, which are difficult to separate. The data displayed in Fig. \ref{chisq} 
are ratios between unnormalized values of $\chi^2$ for the different fit functions. The 
displayed data are for the equilibration time of 25 ms. The other data set has the same
features. For deep potentials, all fits are essentially equally good. At more 
shallow potentials, a Tsallis function reproduces the data better than a 
Gaussian. For the whole range, a double Gaussian gives the best fit. For 
the most shallow potentials, the fitted step becomes too important for $\chi^2$ 
in order to draw any major conclusion from this analysis.

\begin{figure}[h!]
\begin{center}
\includegraphics[scale=0.45]{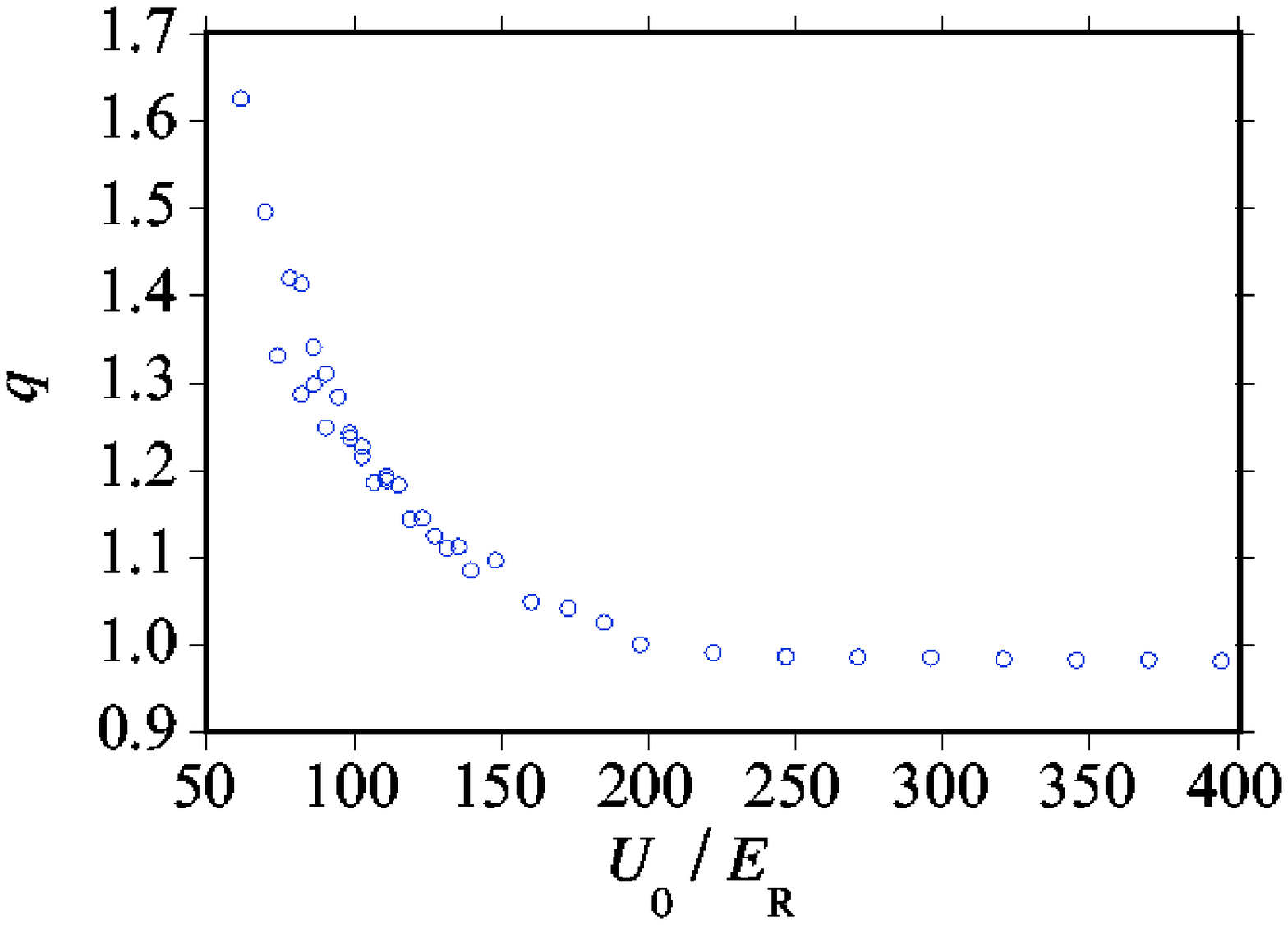}
\caption{(Color online) The fitted Tsallis $q$-parameter as a function of modulation depth 
of the potential for 25 ms equilibration time.
\label{qpara}}
\end{center}
\end{figure}

The parameter $q$ in Eq.~(\ref{eq:tsallis}) can be regarded as a
measure of the shape of the distribution. A $q$ approaching 1 will be
identical to a Gaussian distribution, whereas $q = 2$ corresponds to a
Lorentzian distribution. In Fig.\ref{qpara}, we show a plot of the fitted 
value $q$, for 25 ms equilibration time. For decreasing irradiances, 
$q$ increases smoothly from one, and eventually reaches a value 
higher than $q =1.6$. For the longer equilibration time, the same trend
is evident, but it is much less pronounced, and $q$ dose not reach higher
than $q =1.3$.

\begin{figure}[h!]
\begin{center}
\includegraphics[scale=0.7]{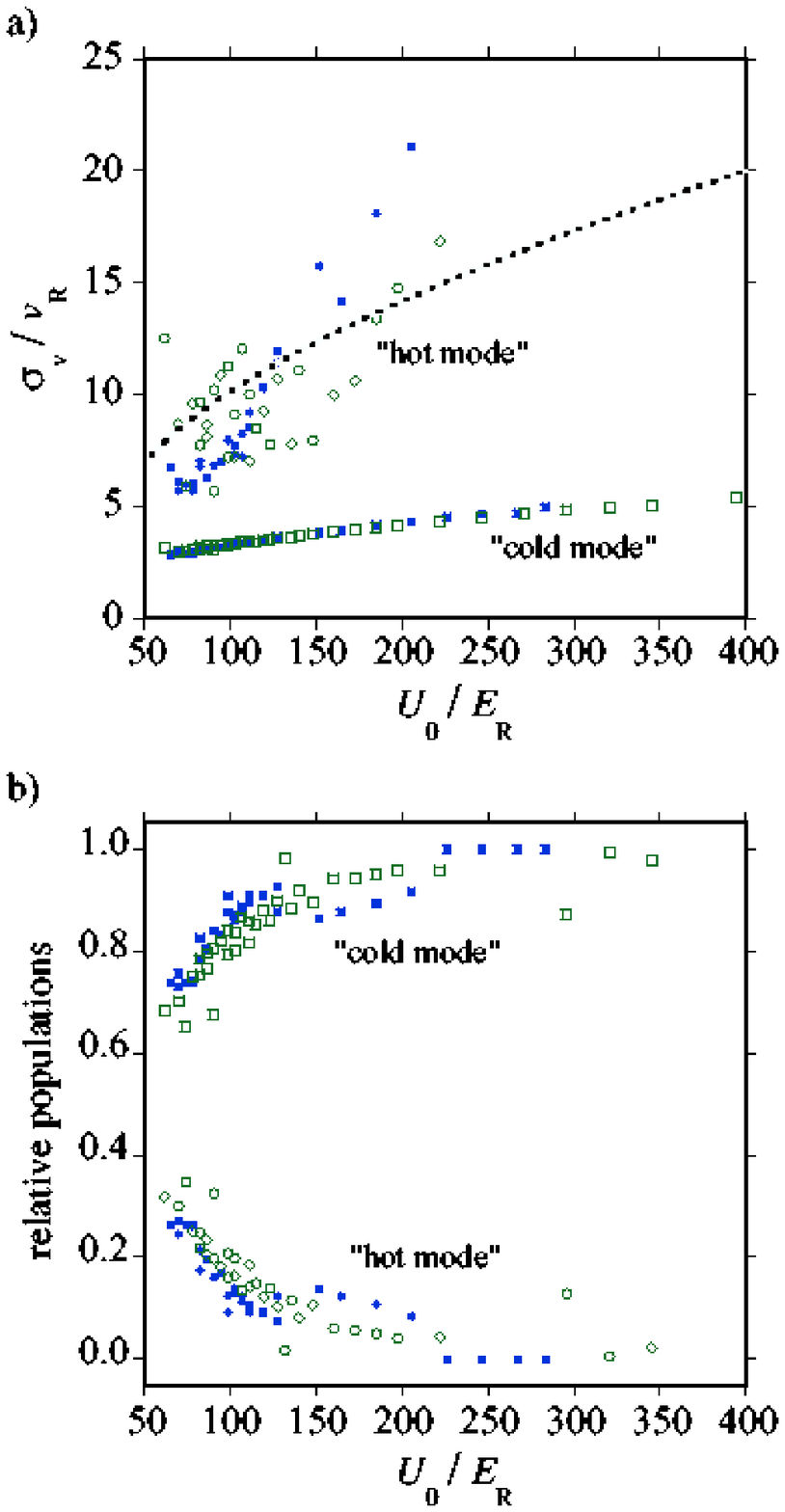}
\caption{(Color online) a) The widths of the two Gaussians as obtained from a fit of the data
to double Gaussians, as a function of modulation depth of the
potential. The dashed line shows the modulation depth in units of velocity. b) The relative population of the two modes of the
population, obtained from the areas under the two Gaussians. 
For both a) and b), filled symbols correspond to data taken with 50 ms
equilibration time and open symbols to 25 ms. Circles are \textquoteleft 
temperatures\textquoteright{} and relative population of the \textquoteleft 
hot mode\textquoteright, and square to the \textquoteleft cold 
mode\textquoteright.
\label{doubleg}}
\end{center}
\end{figure}

The good fit to a double Gaussian can be interpreted as a sign of a 
bimodal velocity distribution. In Fig. \ref{doubleg}a, we show the fitted 
widths of the two Gaussians for both data sets. This should correspond to 
the \textquoteleft temperatures\textquoteright{} of the two modes. Both these \mbox{\textquoteleft temperatures\textquoteright} increase
linearly with potential depths.
The areas of the two Gaussians should be a measure of
the fraction of atoms being in one or the other of the modes.  In Fig. 
\ref{doubleg}b is the calculated relative populations.
The \textquoteleft cold mode\textquoteright{} with 
narrow velocity distribution always contains most of the atoms, but the 
relative number of atoms in the \textquoteleft hot mode\textquoteright{} gets
larger for decreasing potential depths. For potentials deeper than  
$U_{\textrm{0}} = 250 E_{\textrm{R}}$
there is no measurable portion of atoms in the \textquoteleft hot  
modes\textquoteright. The thermal energy of the \textquoteleft hot 
mode\textquoteright{} is of the same order (whithin the large uncertainties) as 
the energy barrier of the optical potential, i.e. the modulation depth (shown in the dashed line in Fig. \ref{doubleg}a).

\section{Numerical simulations}
\label{simulations}
In order to further analyze the results of our experimental data, we
performed numerical simulations for the quantum dynamics of atoms
undergoing Sisyphus cooling. We consider the case of a \eqintext{J=4
\rightarrow J=5} transition, as for the \eqintext{^{133}}Cs atoms used
in the experiments, and for the sake of simplicity we restrict the
motion of the atoms into one dimension (1D). The laser configuration
is the well-known \mbox{1D-lin$\bot$lin} configuration
\cite{Dalibard89} which in fact corresponds to the $z$-direction in
our three dimensional (3D) experimental setup (Fig.~\ref{beamconfig})
with a different lattice spacing. This restriction is legitimate
because, first, the temperature has been shown to be independent of
the lattice spacing \cite{palencia,carminati} and, second, the temperature is
very similar for both 1D- and 3D- configurations (See the comparison
between 3D-experimental and 1D-numerical results in
Fig.~\ref{velspread}. The slight deviation can partly be attributed to the
difference in dimensionality). We first describe the numerical method for the
integration of the dynamics equations (section~\ref{fqmc}) and then we
present the results of the simulations (section~\ref{numerical}).

\subsection{Integration of the dynamics equations}
\label{fqmc}
Consider a two level atom, with Zeeman degeneracy, interacting with a
laser field
\begin{equation}
\overrightarrow{E}_{\textrm{L}} \left( z,t \right ) = \frac{1}{2}
\left \{ \overrightarrow{E}_{\textrm{L}}^+ \left(z\right)\textrm{e}^{-\textrm{i}  
\omega t}
+ \overrightarrow{E}_{\textrm{L}}^- \left(z\right)  
\textrm{e}^{+\textrm{i} \omega t} \right \}.
\label{laser}
\end{equation}
The laser light is strong enough to be treated as a classical field so  
that we can separate the coupling
between the atom and the electromagnetic field into a coupling to the  
laser field and a coupling to vacuum. The coupling to the laser \eqintext{\widehat{V}_{\textrm{AL}}} induces  
a hamiltonian evolution for the atom. On the contrary, because of the coupling to the vacuum modes,
\eqintext{\widehat{V}_{\textrm{AV}}}, the atom is an open quantum  
system for which the evolution has to be treated in
the density matrix formalism. The
evolution of the atom is thus governed by the generalized optical
Bloch equations (OBE) \cite{CCT-JDR-GG88, cohen75}. In the regime of
low saturation, where the experiments are performed, the excited
state relaxes much faster than the typical evolution time
of the ground state and thus it
can be adiabatically eliminated from the OBE. The evolution of the
system then reduces to a master equation for \eqintext{\varrho}, the
atomic density matrix restricted to the ground state
\cite{cohen90}:
\bea
& & \textrm{i} \hbar \frac{\textrm{d}\widehat{\varrho}}{\textrm{d}t} =
          \widehat{H}_{\textrm{eff}} \widehat{\varrho} -
\widehat{\varrho} \widehat{H}_{\textrm{eff}}^\dag
          + \mathcal{L}_{\textrm{relax}} \left( \widehat{\varrho} \right)
\label{master} \\
& & \textrm{where \phantom{aa}}
         \widehat{H}_{\textrm{eff}} = \frac{\widehat{p}^2}{2M} +
\frac{\widehat{V}_{\textrm{AL}}^- \widehat{V}_{\textrm{AL}}^+}
{\hbar\left( \Delta + \textrm{i} \Gamma /2 \right)}.
\label{hamiltonian}
\eea
Here, \eqintext{\widehat{p}} is the momentum operator,  
\eqintext{\Delta} is the detuning between the laser
and the atomic transition, \eqintext{M} is the mass of one atom, and
\eqintext{\widehat{V}_{\textrm{AL}}^\pm =  
-\widehat{\overrightarrow{D}}^\pm
\cdot \overrightarrow{E}_{\textrm{L}}^\pm}
are the raising and lowering parts of the dipole interaction operator
respectively. In Eq.~(\ref{master}),
\eqintext{\widehat{H}_{\textrm{eff}}} is a non-hermitian operator
describing the atom-laser interaction\footnote{
The non-hermitian part of \eqintext{\widehat{H}_{\textrm{eff}}} comes
from the relaxation of the excited state.} 
and \eqintext{\mathcal{L}_{\textrm{relax}}} is an operator
describing the coupling to the vacuum field, i.e. spontaneous
emission of photons. The integration of the master equation is performed
via a full quantum
Monte-Carlo wave function method
\cite{dalibard92,dum92} in which \eqintext{\varrho} is substituted
with a set of stochastic wave functions. The pseudo-hamiltonian 
evolution (first term in Eq.~(\ref{master}))
of each wave function \eqintext{\left| \psi \right\rangle} is governed
by a Schr\"odinger-like equation involving the
non-hermitian hamiltonian \eqintext{\widehat{H}_{\textrm{eff}}}:
\begin{equation}
\textrm{i} \hbar \frac{\textrm{d} \left| \psi
\right\rangle}{\textrm{d}t} = \widehat{H}_{\textrm{eff}} \left| \psi
\right\rangle.
\label{schrodinger}
\end{equation}
  Since equation~(\ref{schrodinger}) does not include the filling terms
  of the ground state from the excited state due to spontaneous
  emission, \eqintext{\left| \psi \right\rangle} is not
  normalized and the instantaneous spontaneous emission rate is given
  by: \eqintext{-\frac{d \langle \psi | \psi \rangle / dt}{\langle \psi
  | \psi \rangle}}. To take into account the emission of photons, the  
pseudo-hamiltonian evolution (Eq. (\ref{schrodinger})) is interrupted by 
 quantum jumps, whose repetition rate is determined with accordance to 
 the spontaneous emission rate. It follows from the emission of a
  photon of wave vector \eqintext{\overrightarrow{\kappa}} and
  polarization \eqintext{\overrightarrow{\epsilon}} that the
  wave function is instantaneously changed into
\begin{equation}
\left| \psi \right\rangle \rightarrow
\left| \psi_{\overrightarrow{\kappa},\overrightarrow{\epsilon}}  
\right\rangle
= \langle 1_{\overrightarrow{\kappa},\overrightarrow{\epsilon}} |
\widehat{V}_{\textrm{AV}} \left( \left| \psi_{\textrm{e}} \right\rangle
\otimes | 0 \rangle \right)
\label{nonhamiltonian}
\end{equation}
with relative probabilities
\eqintext{\left|\left|\psi_{\overrightarrow{\kappa},\overrightarrow{\epsilon}}\right\rangle
\right|^2}. Here the excited state wave function
\eqintext{\left|\psi_{\textrm{e}}\right\rangle =
\frac{\widehat{V}_{\textrm{AL}}^+ \left|\psi\right\rangle}{\hbar
\left(\Delta + \textrm{i} \Gamma /2 \right)}}
is determined by the adiabatic elimination procedure of the excited  
state and \eqintext{| 0 \rangle} and
\eqintext{| 1_{\overrightarrow{\kappa},\overrightarrow{\epsilon}}  
\rangle}
represent the electromagnetic field states respectively without any  
photon, and with one photon of wave vector \eqintext{\overrightarrow{\kappa}}
and polarization \eqintext{\overrightarrow{\epsilon}}. The Monte-Carlo 
 integration then provides a set of time dependent  
stochastic wave functions \eqintext{\left| \psi \right\rangle}, which represent 
the atomic state through the average,
\eqintext{\overline{\sigma}}, of the density matrices associated to the  
wave functions, \eqintext{\sigma = \left|\psi\right\rangle\left\langle\psi\right|}. It is 
then straightforward to show that the quantum master equation for
\eqintext{\overline{\sigma}} is the same as the master equation for the  
actual density matrix
\eqintext{\varrho} (Eq. \ref{master}). Hence, the value of any  
observable \eqintext{\widehat{\mathcal{O}}} for the
quantum system represented by \eqintext{\varrho} is equal to the  
ensemble average of the value of the same
observable for each stochastic wave function represented by  
\eqintext{\left| \psi \right\rangle}
\cite{moelmer93}: at any time,
\begin{equation}
\overline{\left\langle \psi \right| \widehat{\mathcal{O}} \left| \psi  
\right\rangle}
= \textrm{Tr} \left( \widehat{\varrho} \phantom{a}
\widehat{\mathcal{O}} \right).
\label{observable}
\end{equation}

\subsection{Results of the simulations}
\label{numerical}
In this work, we are interested in the particular observable that
represents the momentum distribution of the atoms. We have performed
the full quantum Monte-Carlo integration of the dynamics equations for
a set of \eqintext{200} wave functions for various lattice parameters
(detuning and modulation depth). Because the width of the momentum
distributions are typically broader than several \eqintext{\hbar k},
the spontaneous emission pattern can be approximated by photons
emitted along the 3D coordinate axes $x$, $y$ or $z$. With such an 
approximation, all operators in Eqs.~(\ref{hamiltonian})~and 
~(\ref{nonhamiltonian}) couple states of the form 
\eqintext{\left|m,p\right\rangle} to states of the form 
\eqintext{\left|m',p \pm \hbar k\right\rangle} (where \eqintext{m} and 
\eqintext{m'} represent the internal sub-level of the
atomic ground state).  It is then convenient to perform the
integration in the \eqintext{\left|p\right\rangle}-representation. The 
state \eqintext{\left|\psi\right\rangle} is decomposed onto the
basis of the \eqintext{\left|p\right\rangle} states (with
\eqintext{p=n \hbar k}, with \eqintext{n} an integer positive or
negative). Finally, for usual situations considered in this work, the
typical momenta are smaller than 20 \eqintext{\hbar k}, so that we
take \eqintext{|n| \leq 100}. From the simulations, we determined the
mean kinetic energy as a function of time. After a thermalization
period, the energy reaches a steady-state during which the momentum
distribution was recorded and averaged. The thermalization period was chosen to be \eqintext{1/\Gamma}, corresponding to a time in the order of a millisecond. Since the calculation is performed in 1D, this time cannot be directly compared to the thermalization times in the 3D experiment. 

In order to identify whether the momentum distribution is compatible  
with a Gaussian curve or not, we first compare the root-mean-square
(\mbox{rms}) momentum \eqintext{p_{\textrm{rms}}} defined by  
\eqintext{E_{\textrm{K}}=p_{\textrm{rms}}^2/2M}
(where \eqintext{E_{\textrm{K}}} is the mean kinetic energy of the atomic  
sample) and \eqintext{p_{\textrm{e}}} which represents half
the width at \eqintext{1/\sqrt{\textrm{e}}} of the stationary momentum  
distribution. For a Gaussian distribution, those two values are equal.

\begin{figure}[h!]
\begin{center}
\includegraphics[scale=1.1]{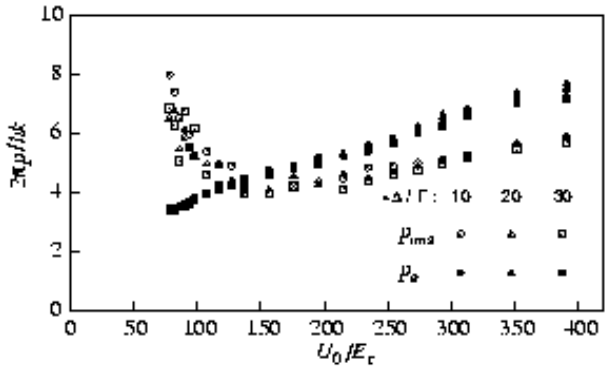}
\caption{Comparison between the rms momentum and the width at  
\eqintext{1/\sqrt{\textrm{e}}} of
the momentum distribution as a function of the potential depth  
\eqintext{U_0} for three different detunings
\eqintext{\Delta=-10\Gamma,-20\Gamma,-30\Gamma}.}
\label{rms}
\end{center}
\end{figure}

\begin{figure}[h!]
\begin{center}
\includegraphics[scale=0.7]{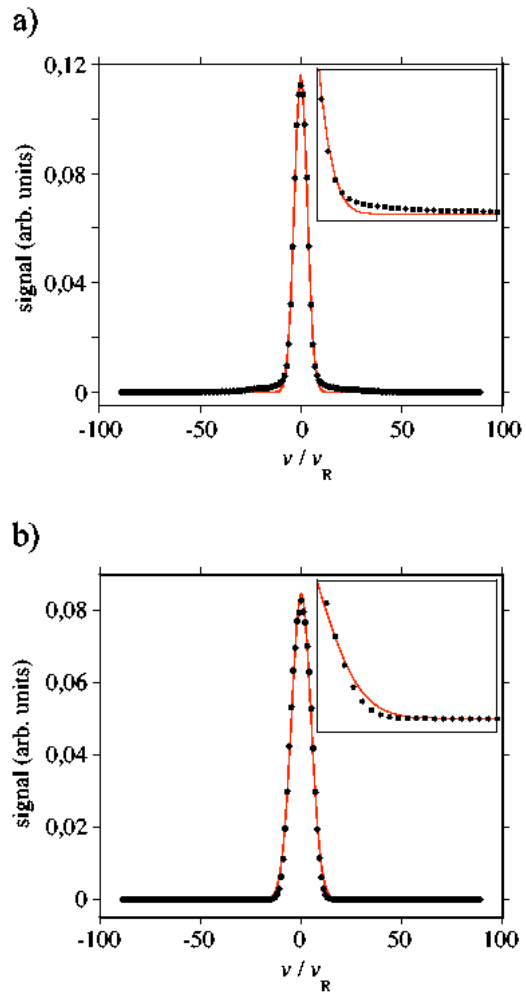}
\caption{(Color online) Numerically recorded velocity distributions with fits to
simple Gaussians. For a, the modulation depth of the  
optical potential was $U_{\textrm{0}}= 78 E_{\textrm{R}}$. 
For b, it was
$U_{\textrm{0}} = 235 E_{\textrm{R}}$. The insets show magnifications
of portions of the wings of the distributions.
\label{numveldistr}}
\end{center}
\end{figure}

We plot in Fig.~\ref{rms}, the numerical results for  
\eqintext{p_{\textrm{rms}}} and \eqintext{p_{\textrm{e}}} as a function of  
the potential depth \eqintext{U_0} for three different detunings  
\eqintext{\Delta}.
We find that these values are independent of the detuning within the
numerical errors. Several points for lower values of \eqintext{U_0}
have also been calculated but the atomic cloud was found not to
thermalize. For those cases, the temperature increases more or less linearly and the velocity
distribution becomes almost flat. It is also clear in Fig.~\ref{rms} that
\eqintext{p_{\textrm{rms}}} and \eqintext{p_{\textrm{e}}} have
different behaviors.  \eqintext{p_{\textrm{rms}}} reproduces the well
known dependence of the kinetic energy versus the modulation depth:
\eqintext{p_{\textrm{rms}}} scales as \eqintext{\sqrt{U_0}} 
for high values of \eqintext{U_0} and abruptly increases as
\eqintext{U_0} reaches very low values, typically lower than 150
\eqintext{E_{\textrm{R}}} (the point of d\'ecrochage). The minimum
value of \eqintext{p_{\textrm{rms}}} is found to be of the order of
\eqintext{\left(p_{\textrm{rms}}\right)_{\textrm{min}} \simeq 4.1
\hbar k}. On the contrary, we find that \eqintext{p_{\textrm{e}}}
increases monotonically versus \eqintext{U_0} for low values as well
as for high values of \eqintext{U_0}. The minimum value of
\eqintext{p_{\textrm{e}}} is obtained for the minimum value of
\eqintext{U_0} for which a steady-state velocity distribution can be
obtained (\eqintext{U_0 \gtrsim 78 E_{\textrm{R}}}) and is found to be
of the order of \eqintext{\left(p_{\textrm{e}}\right)_{\textrm{min}}
\simeq 3.4 \hbar k}. We identify two different regimes that can be
distinguished: For \eqintext{U_0} above d\'ecrochage (\eqintext{U_0
\gtrsim 150 E_\textrm{R}}), both \eqintext{p_\textrm{e}} and \eqintext{p_\textrm{rms}} increase
and \eqintext{p_\textrm{e}} is slightly larger than \eqintext{p_\textrm{rms}} that is
to say that the momentum distribution is wider than a Gaussian
distribution with the same \eqintext{p_\textrm{rms}}. For \eqintext{U_0}
below d\'ecrochage (\eqintext{U_0 \lesssim 150 E_{\textrm{R}}}),
\eqintext{p_{\textrm{e}}} decreases while \eqintext{p_{\textrm{rms}}}
increases rapidly as \eqintext{U_0} decreases; the momentum
distribution has large wings and becomes narrower than a Gaussian
distribution. These different characteristics are illustrated in
Fig.~\ref{numveldistr} where we plot the simulated velocity
distributions together with Gaussian fits in the two regimes
\eqintext{U_0 \lesssim 150 E_{\textrm{R}}} and \eqintext{U_0 \gtrsim
150 E_ {\textrm{R}}}.

One should note that this result is in disagreement with earlier
calculations performed for atoms with a theoretical \eqintext{J=1/2
\rightarrow J=3/2} transition for which Castin et al. find that 
$p_{\textrm{rms}}>p_{\textrm{e}}$ for any value of the potential depth
$U_{0}$ \cite{Castin}. In fact, when running
our simulation for the \eqintext{J=1/2 \rightarrow J=3/2} transition,
we were able to reproduce the results of \cite{Castin} and we thus
conclude that the discrepancy is due to the different atomic
transitions considered in \cite{Castin} and in the present work. We
finally conclude that in general, the momentum distribution
significantly differs from a Gaussian distribution. Moreover, we find
that the threshold for \eqintext{p_{\textrm{rms}}} at low values of
\eqintext{U_0} do not affect \eqintext{p_{\textrm{e}}}.

\begin{figure}[h!]
\begin{center}
\includegraphics[scale=0.4]{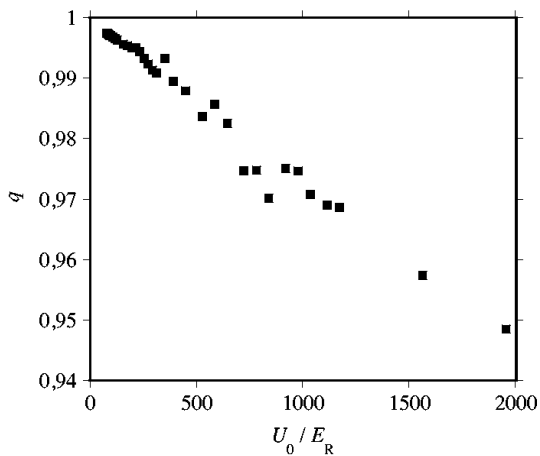}
\caption{
The $q$-parameter as a function of modulation depth of the
potential, obtained from fitting the numerically computed data to
Tsallis functions.
\label{numqpara}}
\end{center}
\end{figure}

We now turn to a more detailed analysis of the momentum distributions. 
We first fit the velocity distributions to Tsallis functions. The
dependence of the Tsallis parameter \eqintext{q} on the modulation
depth is also shown in Fig.~\ref{numqpara} and show a linear dependence
of \eqintext{q} versus \eqintext{U_0}.

For all numerical data \eqintext{q} differs from \eqintext{1} only by
less than \eqintext{5~\%} and is less than \eqintext{1}. Moreover,
\eqintext{q} is found to tend to \eqintext{1} for shallow potentials
indicating that the best Tsallis fit is close to a Gaussian curve in
disagreement with the previous discussion.  
The discrepancy between numerical simulations and experimental
measurements may be caused by the different dimensionality considered
in the experiments and in the simulations.

\begin{figure}[h!]
\begin{center}
\includegraphics[scale=0.55]{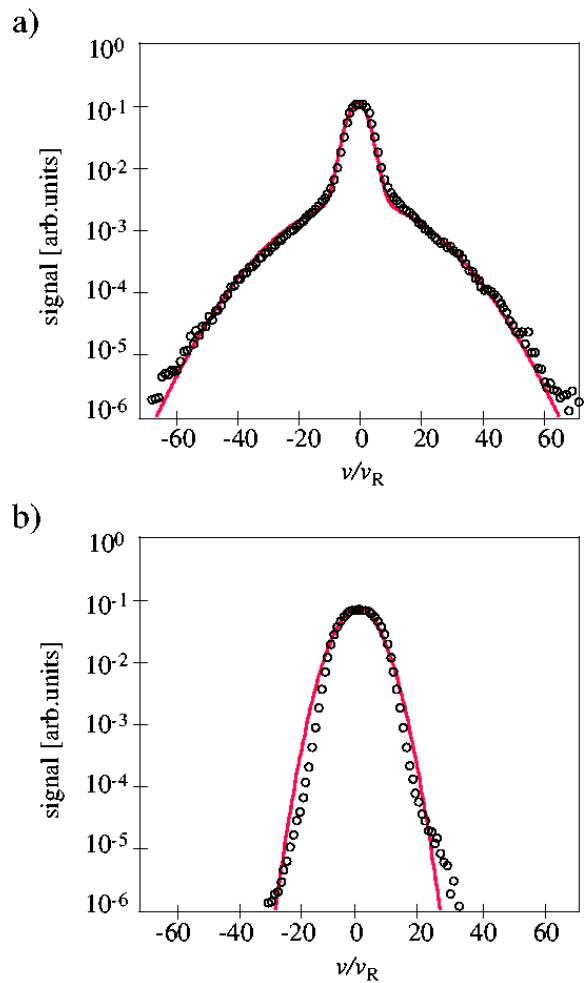}
\caption{(Color online) 
Numerically computed velocity distribution for
a) $U_{\textrm{0}} = 78 E_{\textrm{R}}$ and b) $U_{\textrm{0}}= 293  
E_{\textrm{R}}$ together with fits to Double Gaussians.
\label{numdglog}}
\end{center}
\end{figure}

Consider now fits to double Gaussians. We plot in Fig.~\ref{numdglog}
numerically recorded velocity distributions in logarithmic scale for
potential depths in the two regimes corresponding to shallow and deep
potentials, together with fits to double Gaussians. For the deep
potential (\eqintext{U_0 = 293 E_{\textrm{R}}}), the profile is
essentially parabolic and thus corresponds to a Gaussian distribution. 
For the shallow potential (\eqintext{U_0 = 78 E_{\textrm{R}}}), we
clearly identify two contributions: in addition to a narrow parabolic
profile (corresponding to low energetic atoms), a broad one
(corresponding to high energetic atoms) appears.

\begin{figure}[h!]
\begin{center}
\includegraphics[scale=0.8]{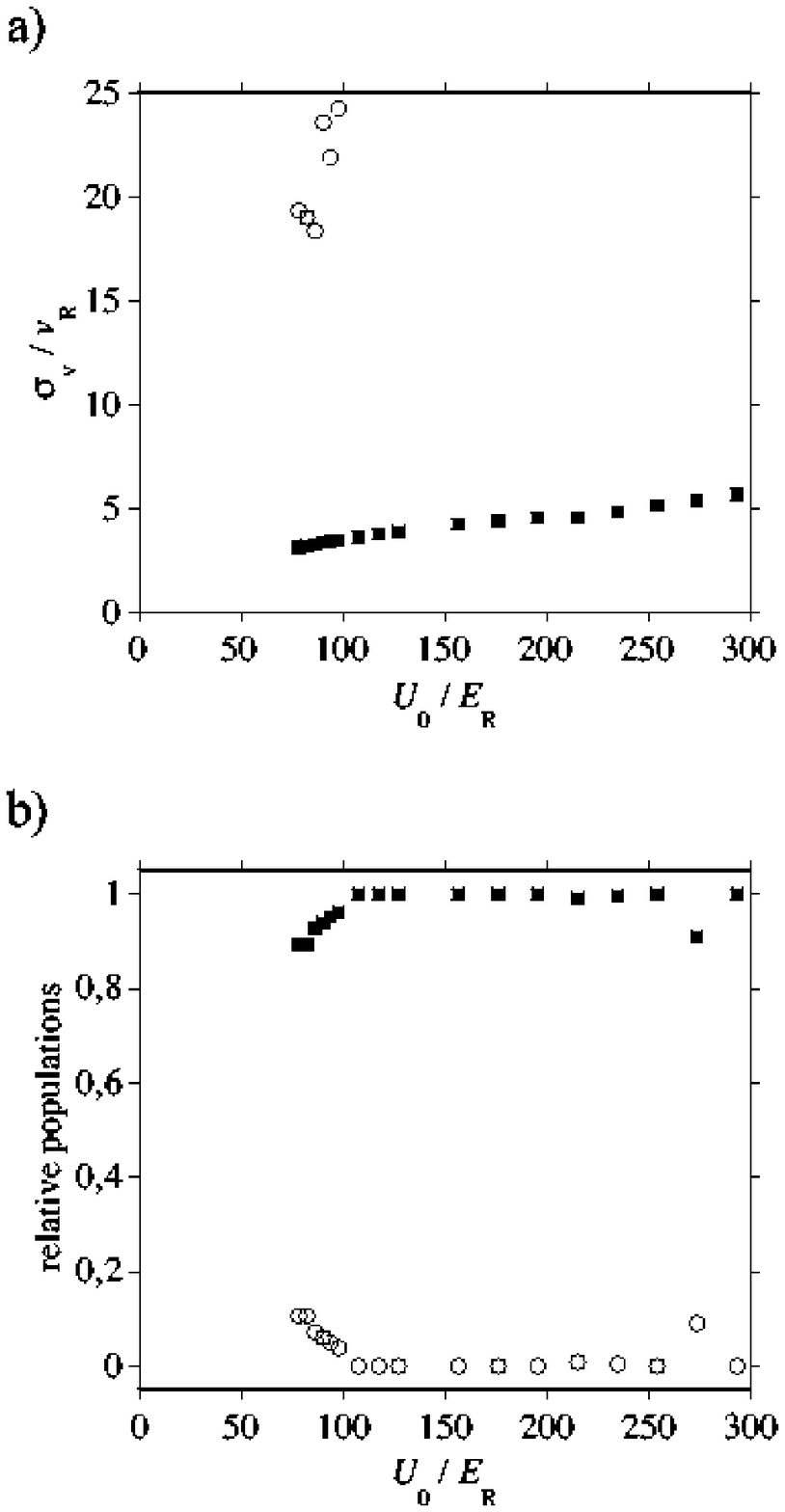}
\caption{
a) Widths of the two Gaussians (cold mode: squares, hot mode:
circles) as obtained from
a fit of the numerical data to Double Gaussians, as a function of  
modulation depth of the potential.
b) The relative population of the two modes of the velocity  
distributions, obtained from the areas
under the two Gaussians. The cold mode (squares)  
corresponds to the large fraction whereas
the hot mode (circles) corresponds to the small fraction.
\label{numdoubleg}}
\end{center}
\end{figure}

This supports the interpretation of the dynamics in terms of a bimodal  
atomic distribution, with each mode corresponding
to \textquoteleft trapped\textquoteright{} atoms and to nearly \textquoteleft 
free\textquoteright{} atoms. The whole distribution is 
well fitted by a double Gaussian function. We plot in
Fig.~\ref{numdoubleg}a the widths of both the modes as functions of  
\eqintext{U_0}.
The numerical results are in good agreement with experimental ones
(see Fig. \ref{doubleg}). For shallow potentials, we find two Gaussian components with
widths that both increase with the potential depth $U_0$, whereas for deep
potentials, the \textquoteleft hot component\textquoteright{} is almost undetectable.Thus the route to \textquoteleft d\'ecrochage\textquoteright{} for shallower 
potentials  can be interpreted as a transfer from the cold mode to
the hot mode. This is supported by the results for the populations of  
the two Gaussian contributions to the velocity
distribution plotted in Fig.~\ref{numdoubleg}b. We actually find  
that the cold mode is largely dominant
even for very shallow potentials close to \textquoteleft 
d\'ecrochage\textquoteright.

Finally, we compare the numerical and experimental results.
A direct quantitative comparison is not adequate, since the simulations  
are done in 1D. However,
qualitatively, the experimental data are reproduced excellently. Figures 
~\ref{numveldistr}-\ref{numdoubleg} show numerical data  
corresponding to the experimental ones
in Figs.~\ref{veldistr25}-\ref{doubleg}. The single Gaussian works for high 
irradiance but fails to fit the  
wings of the distribution for low irradiances. A Tsallis function does not 
fit the distribution any better than a  
single Gaussian for the numerical data. Again, the distribution is best 
fitted by a double Gaussian and
this is particularly pronounced for shallow potentials. The fits to double 
Gaussians also reproduce the signature of one \textquoteleft hot\textquoteright{} 
and one \textquoteleft cold\textquoteright{} mode for shallow potentials.
This strongly supports assumptions of a bimodal distribution.

We find that the `cold' mode is largely dominant even for very shallow potentials close to \textquoteleft d\'ecrochage\textquoteright in both the experimental
($70\ \%$) and the numerical ($90\ \%$) results (see Fig.~\ref{doubleg}b and \ref{numdoubleg}b). The quantitative discrepancy between the limit population of the `hot' and
`cold' modes results from the different dimensionnalities (3D experiments versus 1D calculations). Indeed, the fraction of bound atoms in a $d$D dimension situation can
be estimated by
\be
n_{\textrm{bound}}^{(d\textrm{D})} = \alpha \sum_{j}{p_{d\textrm{D}}(E_{j})} \prod_{1 \le \mu \le d}{\mathbf{1}_{E_{j \mu} < E_{\textrm{max} \mu}}}
\label{nlie1}
\ee
where $p_{d\textrm{D}}(E_{j})$ represents the population of a state of energy $E_{j}$, $\alpha$ in a normalization factor and $E_{\textrm{max} \mu}$
stands for the maximum energy of bound states in the potential wells along the direction $\mu$ ($E_{\textrm{max} \mu}$ is of the order of the potential depth).
Now, if the space directions are separated (which is the case in the harmonic approximation that one can assume to be valid for bound states), then
\be
n_\textrm{bound}^{(d\textrm{D})} = \prod_{1 \le \mu \le d}{\alpha_\mu \sum_{j}{p_{1\textrm{D}}(E_{j \mu})} \mathbf{1}_{E_{j \mu} < E_{\textrm{max} \mu}}} ~.
\label{nlie2}
\ee
Hence, assuming that all directions are equivalent, the fraction of atoms in non-bound states reads
\bea
n_\textrm{non bound}^{(d\textrm{D})} & = & 1 - \prod_{1 \le \mu \le d}{\left( 1 -n_\textrm{non bound}^{(1\textrm{D} \mu)} \right)} \nonumber \\
& \sim & d n_\textrm{non bound}^{(1\textrm{D})} ~.
\label{nlie3}
\eea
Therefore, the limit populations of the `hot' mode at `d\'ecrochage' are consistent in the experiments ($30\ \%$ and $d=3$) and in the simulations ($10\ \%$ and $d=1$).

\section{Conclusions}
In this work, we have studied the velocity distributions of cold atomic
samples obtained by Sisyphus cooling both in experiments with
\eqintext{^{133}}Cs and in full quantum numerical simulations
performed for the actual \eqintext{4 \rightarrow 5} transition of
\eqintext{^{133}}Cs. We stressed in particular the deviation from a
Gaussian distribution. This has already been forecasted via
semi-classical as well as quantum simulations for a simplified
\eqintext{1/2 \rightarrow 3/2} transition showing the difference of
the rms velocity \eqintext{v_{\textrm{rms}}} and the velocity
\eqintext{v_{\textrm{e}}} corresponding to half the width at
\eqintext{1/\sqrt{\textrm{e}}} of the distribution \cite{Castin}. We
recovered such a property but with a significantly different behavior
of the ratio \eqintext{v_{\textrm{rms}}/v_{\textrm{e}}}. This shows
that the non-Gaussian behavior of the velocity distributions is
certainly not a trivial effect in Sisyphus cooling.

\subsection{Summary of our results}
Our results (experiments as well as numerical simulations) show that the velocity distributions are compatible with Gaussian functions for deep enough potentials (typically for $U_0$ larger than a hundred recoil energies). Note that in this case, the atoms are trapped in the potential wells (i.e. the kinetic energy of the atomic cloud is significantly smaller than the potential depth). The deviation of the velocity distribution from Gaussian functions become more prominent for shallow light shift potentials. We tested several types of functions to better fit the shape of the velocity distributions in the range of parameters corresponding to deeper potentials. We found that a better fit (corresponding to smaller $\chi^2$) can be obtained by using a Tsallis function or a double Gaussian.

\paragraph*{Tsallis functions -}
The use of a Tsallis function is related to the details of the dynamics of atoms cooled by the Sisyphus mechanism which is known to be slightly more complicate than a Brownian motion. The Tsallis function introduces a new parameter $q$ which deviation from $1$ measures the deviation of the velocity distribution from a Gaussian function. The parameter $q$ can be calculated in the `jumping regime' \cite{mennerat98} and it is straightforward to show that $q$ tends to $1$ for high values of the potential depth (thus corresponding to a weak deviation from a Gaussian), and increase for shallow potentials. An {\it ab initio} calculation of $q$ is more tricky in the `oscillating regime' which correspond to the domain of parameters for shallow potentials, near the point of `d\'ecrochage' \cite{mennerat98}. Nevertheless we can plot the value of $q$ corresponding to the best fit of the measured velocity distributions as a function of modulation depth. For large modulation depths, we find that $q$ approaches $1$, which corresponds to a Gaussian distribution, in agreement with the analytical calculation (see section \ref{Fitting functions and motivations} and \cite{Lutz}). When reducing the potential depth, we clearly observed an increase in $q$ and this corresponds to a velocity distrubution with wings larger than in a Gaussian function. In our case the maximum $q$ is close to 1.6 and this corresponds in our experiments to a potential depth $U_0 \simeq 60 E_\textrm{R}$. For $U_0 < 60 E_\textrm{R}$, the atomic cloud does not reach a steady state and the optical lattice disintegrates. It is interesting to note that the rms velocity of Tsallis distributions with $q$ above $q_{\textrm{cr}}$ = 5/3 diverge \cite{Tsallis99}. If one would plot rms `temperatures' of the atoms using the rms velocity, this would correspond to a diverging temperature. As one is often limited by noise in the wings of the velocity distribution, one has a tendency to restrict the analysis to atoms with velocities several times below the 1/{\textrm{e}} value of the distribution. Any divergence is hence avoided. Note also that such divergences are very familiar: the wings of a Lorentz distribution are also known to cause a divergence of the rms value of the distribution. One can also recall that in the case of narrow line cooling, the rms velocity diverges \cite{Wallis89, Katori99} when one approaches the atomic resonance, and that for very small detunings one can no longer even have a normalized distribution function \cite{Wallis89}.

\paragraph*{Double Gaussian functions -}
Fitting the recorded (experimental or numerical) distribution functions to double Gaussians works even better than the Tsallis function. On the one hand, it is not surprising that a fitting procedure with more free parameters gives better fits. On the other hand, the velocity distribution in logarithmic scale in Fig. \ref{numdglog}a clearly exhibits two components with very different widths. For deep potentials, one recovers a Gaussian distribution of `cold atoms' bound in in the potential wells as expected from the above discussion. When decreasing the potential depth, a fraction of `hot atoms' grows up (for $U_0 < 120 E_\textrm{R}$). These atoms have an energy larger than the potential depth and are not trapped in the potential wells. We found that the fraction of `hot atoms' can be significant for very shallow potentials. It reaches $30\ \%$ in 3D experiments and $10\ \%$ in 1D numerical simulations just above `d\'ecrochage' at $U_0 \simeq 60 E_\textrm{R}$ (the discrepancy between the experiments and the simulations is due to the different dimensions as shown at the end of section~\ref{numerical}). This result strongly supports assumptions that an optical lattice has a {\it bimodal velocity distribution}. A straightforward interpretation would be that some atoms are bound at lattice sites, whereas others have enough energy to move around on top of the modulated potential. An interesting results of our work is that the phenomenon of `d\'ecrochage' does not correspond to a sharp increase of the width of the velocity distributions corresponding to each mode but to a continuous transfer from the `cold mode' to the `hot mode'. We found that when d\'ecrochage occurs, the fraction of atoms in the `hot mode' does not exceed a few tens percent.

\subsection{Perspectives}

The results shown in this paper stronlgy suggest that the simple picture for Sisyphus cooling, based on a competition between a diffusion and a friction (see Eq.~\ref{eq:sisyphus_nonlinear}),
is not adequate to describe the \textquoteleft coldest\textquoteright{} velocity distributions. Even though one has to be careful before generalizing the conclusions of this paper to other situations of laser cooling and/or trapping, the existence of two velocity modes might provide a useful guide to understand the dynamics and limits of laser cooling. One can note e.g. that for shallow potentials, one has fewer bound states, and the fraction of atoms in the conduction band gets more prominent, as shown in Figs.~\ref{doubleg} and \ref{numdoubleg}.

These atoms will experience a friction force corresponding to the classical Sisyphus cooling model. The route to equilibrium for the bound atoms is less clear. One hypothesis \cite{castpriv} is that bound levels are uniformly \textquoteleft watered\textquoteright{} from the conduction band, whereas high lying levels are more likely to escape. Thus, the route to equilibrium is not quite a competition between cooling and heating. A drawback with this theory is that it would not yield Gaussian velocity distributions. However, this theory has the advantage that the rate of equilibration should depend linearly on irradiance, which is consistent with previous experiments \cite{raithel, palencia2}. In contrast, the standard Sisyphus cooling theory predicts a cooling rate independent of irradiance \cite{Dalibard89}. An interesting experiment would be to measure the velocity distribution as a function of time after a sudden change of the light shift potential, and see if the two populations would evolve differently.

Figures \ref{veldistr25} and \ref{veldistr50} seem to indicate a time dependence
of the experimentally recorded velocity distribution. However, with the current data set (using only
the two cooling times 25 ms and 50 ms), and with the current experimental uncertainties, we cannot draw any quantitative conclusion for the time dependence of the 
velocity distributrion. In future work, we
will study the velocity distribution as a function of time.

It would also be interesting to extend the test functions used in
this paper to a narrow-line cooling scheme, which become more and
more used with the laser cooling of earth-alkaline atoms. At this
stage, one can however note, that a non normalized distribution
function will have as an effect that there is no steady state
distribution and that in this case atoms will diffuse to large
velocities.\ This will appear in an experiment as a leakage rate of
the atoms from the optical lattice. The background observed in our
experiment become more and more dominant for shallow potential wells.
One might expect this to have a contribution from a diffusion of the
atoms beyond the capture range of the optical lattice corresponding
in practice to a non-normalized distribution function. A detailed
analysis of the velocity distribution of atoms in optical lattices
thus appears as a promising tool to study new statistical effects.

Experiments as well as full quantum simulations (in 1D and 3D) should allow one to
get new insights in the dynamics of such systems. Apart from the
suggestions above, future work could e.g. focus on the phase space
dynamics of atoms in optical lattices and of quantum transport
properties of ultra-cold atoms or even Bose-Einstein condensates.
\\

\begin{acknowledgments}
LSP thanks the swedish group for warm hospitality during the period
when a part of this work was achieved.  He also acknowledges financial
support from the Swedish Foundation for International Cooperation in
Research and Higher Education (STINT).  RK thanks the Wenner-Gren  
foundation for a travel grant. We would like to thank Dr.
Peter Olsson at Ume{\aa} University for letting us use the LINUX
cluster and also for support during the simulations at the theoretical
physics department at Ume{\aa} University. We also thank Eric Lutz and Philippe Verkerk for fruitful discussions.

This work has been supported by the Carl Trygger foundation and the  
Knut \& Alice Wallenberg foundation.
\end{acknowledgments}

\end{document}